\documentstyle[aps,epsf]{revtex}

\begin{document}
\title{Geometry-induced smoothing of van Hove singularities in capped carbon nanotubes}
\author{D.V. Kolesnikov$^{1}$ and V.A. Osipov$^{2}$}
\address{Joint
Institute for Nuclear Research, Bogoliubov Laboratory of
Theoretical Physics, 141980 Dubna, Moscow region, Russia\\
e-mail: $^1$kolesnik@theor.jinr.ru, $^2$osipov@theor.jinr.ru}
\date{22 December, 2006}
 \maketitle

\begin{abstract} The electronic states of capped semi-infinite
nanotubes are studied within the phenomenological gauge
field-theory model. A single manifold for the description of both
the nanotube and the cap region (considered as nearly a half of
either Ih or I fullerene) is suggested. The wavefunctions and the
density of states (DoS) are numerically calculated for both
metallic and semiconducting nanotubes. The smoothing of van Hove
singularities is found and proven analytically. The comparison
with the experimental observations is discussed.
\end{abstract}
\bigskip The electronic properties of capped carbon nanotubes are
essential to proposed device applications. The localized
eigenstates in a cap region were predicted theoretically
in~\cite{tamura}. Experimentally, the signature of resonance
states localized at the ending caps of a nanotube was observed in
the spectroscopic measurements on carbon nanotube
caps~\cite{carroll}. Another spectroscopic analysis of carbon caps
performed in~\cite{kim} shows spatially resolved electronic
properties along capped nanotubes and highlights the van Hove
singularities (VHS) in nanotube bulk and resonance states at the
end. The scanning tunneling spectroscopy (STS) gives information
about the local density of states (LDoS). Several sharp peaks
experimentally observed in~\cite{kim} were attributed to the VHS.
The tight-binding calculations show good agreement with the
experimental data (see, e.g.,~\cite{kim,rubio}).

The aim of our paper is twofold. First, we suggest an appropriate
geometry for the shape of capped nanotubes and develop a continuum
field-theory model to describe the electronic states. Second, we perform
numerical calculations to obtain the structure of DoS. This study gives
us an important information about the influence of the cap region on the electronic
properties of the nanotube and illuminates the behavior of the VHS.
We will consider a closed semi-infinite nanotube with a spherical
cap. It is suggested that the nanotube region is defect free
(contains only hexagons) and infinitely long, so that
edge effects are negligible. The cap is a half of the spherical (I or Ih)
fullerene and contains exactly six pentagons. A continuum field-theory
model for the description of low-energy electronic
states in carbon nanostructures of arbitrary geometry was formulated
in~\cite{kolesn_epjb} in the form
\begin{equation}
-i\sigma^{\alpha}e_{\alpha}^{\ \mu}(\nabla_{\mu} -
ia_{\mu}^k-iW_{\mu})\psi^k = E\psi^k. \label{eq:5a}
\end{equation}
Here $\sigma^\mu$ are the conventional Pauli matrices ($\mu=1,2$),
the energy $E$ is accounted from the Fermi energy, the Fermi
velocity $V_F$ is taken to be one, and the two-component wave
function $\psi$ represents two graphite sublattices ($A$ and $B$).
By using the index $k$ in (\ref{eq:5a}) we take into account
electronic states at two independent Fermi wave vectors in carbon
lattice (so-called "K-spin up" ($K^{\uparrow}$) and "K-spin down"
($K^{\downarrow}$) states). Geometry is involved via zweibeins
$e_\mu^\alpha$, \ $\nabla_{\mu}=\partial_{\mu}+\Omega_{\mu}$ where
the spin connection term $\Omega_{\mu}=(1/8)\omega^{\alpha\
\beta}_{\ \mu} [\sigma_{\alpha},\sigma_{\beta}]$, and
$(\omega^{\mu})^{\alpha\beta}$ are the spin connection
coefficients:
\begin{equation}
(\omega_\mu)^{ab}=e^a_\nu (\partial_\mu e^{\nu
b}-\Gamma^{\nu}_{\mu\chi}e^{\chi b})=-(\omega_\mu)^{ba},
\end{equation}
where
\begin{equation}
\Gamma^{\nu}_{\mu\lambda}=\frac{g^{\nu\chi}}{2}(\partial_{\mu}g_{\chi\lambda}+
\partial_{\lambda}g_{\chi\mu}-\partial_{\chi}g_{\mu\lambda})
\end{equation}
are the metric connection coefficients.

In order to take into account pentagons in the cap region, two
compensating gauge fields $a_{\mu}$ and $W_{\mu}$ are introduced
in (\ref{eq:5a}). The Abelian field $W_\mu$ is responsible for the
elastic flux due to pentagonal defect~\cite{osipov_pla,jpa99}. Its
circulation around a single disclination is found to be exactly
$2\pi/6$ (the appearance of a pentagon in the honeycomb lattice is
equivalent to the creation of $60^{\circ}$ disclination). The
non-Abelian field ${\vec a}$ allows us to take into account the
exchange between A and B sublattices in the presence of a
pentagon~\cite{jose93,crespi,kolesn_epjb}. Namely, this exchange
can be described by using an appropriate boundary condition for
the K spin part of the four-component spinor wavefunction
$\psi=(\psi^\uparrow\psi^\downarrow)^T$ in the form
$\psi(\varphi+2\pi)=-T\psi(\varphi)$. Here the holonomy operator
$T$ is composed as a product of two operators in the form
$\exp(i\Phi\tau_i)$, $i$=2,3  where the isospin Pauli matrices
$\tau$ act on the K part of the spinor components. The general
consideration takes proper account of the relative placement of
pentagons. For even number of defects one has (see~\cite{crespi})
\begin{equation}
\oint a^k_{\mu} dx^{\mu}=\pm
(N\frac{2\pi}{4}+M\frac{2\pi}{3}).\label{acirc2}
\end{equation}
Here the sign plus (minus) is taken for $k=K^{\uparrow}$
($k=K^{\downarrow}$), respectively, N is a number of defects and M
($M=n-m$ (mod 3)) depends on the arrangement of pentagons, n and
m are the numbers of steps in positive and negative directions,
respectively. The directions rotated by $2\pi/3$ are considered to
be identical. For (I) fullerenes $M=M_2=\pm 1$ or 0 for any two
defects, $M_4=0$ for any four defects,  and $M_6=-M_2$ for six
defects. In the case of (Ih) fullerenes M is always equal to zero
due to the mirror symmetry of the lattice
(see~\cite{kolesn_epjb}).

\begin{figure}[ht]
\begin{center}
\epsfysize=4cm\epsffile{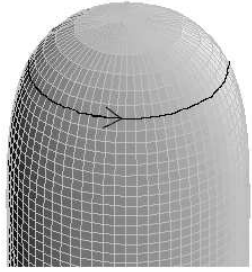} \caption[]{The modelled
geometry of closed tube. A contour $\Gamma_S$ encircles the
surface $S$. The z-axis is directed downside.} \label{fig1}
\end{center}
\end{figure}

In the isospin space, the action of the operator T corresponds to
a rotation by $2\pi N/4$ around the second axis and then by $2\pi
M/3$ around the third axis. The angle between the initial and
final vectors gives exactly the phase (\ref{acirc2}). Notice that
for odd number of defects the first transformation places the vector
on the third axis, so that the second transformation does not change the angle
(phase). To simplify a problem, let us introduce an effective
field for uniformly distributed defects instead of six point-like
fluxes due to pentagons in such a way that the phase factors in
(\ref{acirc2}) remains the same. Let $\Gamma_S$ be the circle on
the (hemi)sphere (with the unit radius, for simplicity) which
encircles the area S including the pole (see fig.~\ref{fig1}).
Evidently, $0<S\leq 2\pi$.
\bigskip
It is easy to find that the operator of rotation
around the second axis in the isospace takes the form $\exp(i\tau_2 S
3/2)$.  Indeed,  we have six defects (each rotating on $2\pi/4$) continuously
distributed on the area $2\pi$. The rotation angle is
$\Phi_M=3S/2$.
The operator of rotation
around the third axis can be
approximated as $M(S)=-M_2 \cos(3(S-2\pi)/4)$.  This agrees
with the properties of M described above: $M(2\pi/3)=M_2$,
$M(4\pi/3)=0$, $M(2\pi)=-M_2$. The rotation angle is written as
\begin{equation}
\Phi_T=-\frac{2\pi}{3} M_2 \cos(3(S-2\pi)/4).
\end{equation}
The total operator $T$ in the form
\begin{equation}
T=\exp(i\tau_3 \Phi_T)\exp(i\tau_2\Phi_M)
\end{equation}
corresponds to the phase rotation by angle $\Phi$ which is
determined from the equation
\begin{equation}
\cos\Phi(S)=\cos\Phi_M(S) \cos\Phi_T(S).
\end{equation}
As a final step, the circulation of the field $\vec a$ along
$\Gamma_S$ is
\begin{equation}
\oint_{\Gamma_S}a_\mu\; dx^\mu=\pm \Phi(S).
\end{equation}
On the equator of the sphere the circulation is
$\Phi(2\pi)=3\pi+2\pi M_2/3$. Since the equator is connected with
the tube region, any contour around the tube encircles the same
quantity of the field sources as $\Gamma_{2\pi}$. Therefore,
finally in the spherical region and on the tube we obtain the
following conditions:
\begin{eqnarray}
\oint_{\Gamma_S} a_\varphi d \varphi=\pm
\Phi(S),\label{sfield}\\
\int_0^{2\pi} a_\varphi d \varphi=\pm (3\pi-2\pi
M_2/3).\label{fields}
\end{eqnarray}
Notice that the factor $3\pi$ in (\ref{fields}) can be excluded by
the redefinition of the momentum $j$ (see below), so that the
circulation in the tube region (\ref{fields}) is determined by $M_2$
in agreement with~\cite{meletube}.
For both (Ih) fullerenes and (I) fullerenes
with $M_2=0$ one has $\Phi=\Phi_M=3S/2$, which agrees with
monopole approximation \cite{jose92,kolesn_epjb}. Therefore, one
can conclude that metallic tubes can be capped by either (Ih)
fullerenes (armchair) or (I) fullerenes with $M_2=0$ while the
semiconducting tubes are capped only by (I) fullerenes with
$M_2\neq 0$.

In the spherical region, the  circulation of the field $W_\varphi$
has the standard form similar to~\cite{kolesn_epjb}
$$
\oint_{\Gamma_S} W_\mu dx^\mu=-S,
$$
whereas in the tube region its circulation is equal to $-2\pi$ and may be ignored.


The continuum description in (\ref{eq:5a}) implies a single manifold
with the curvature being a continuous smooth function.
Therefore we have to propose appropriate geometry for the capped tube:
a single manifold $\Sigma$ which reproduces the sphere in the cap region
and the tube outside. Let us use the following manifold:
\begin{eqnarray}
\vec R (\rho(z) \cos\varphi,\rho(z)\sin\varphi,z),\;
  \rho(z)=R_t \sqrt{1-\exp(-2\Lambda)},
\end{eqnarray}
with $$ \Lambda=\frac{z+R_f}{R_f},\, \alpha=R_t/R_f,\, z\geq
-R_f,\,
    0\leq\varphi<2\pi,$$ where $R_f$ characterizes the capped
area and $R_t$ is the tube radius. It is taken into account that
$R_f$ can generally be different from $R_t$. To simplify the
problem, one can approximate $S=2\pi\Lambda$ as if the surface would
be the hemisphere with the radius $R_f$. The metric tensor is written
as
\begin{equation}
g_{zz}=\alpha^2\frac{e^{-4\Lambda}}{1-e^{-2\Lambda}}+1,\quad
g_{\varphi\varphi}=\rho^2(z),\quad
g_{z\varphi}=0.
\end{equation}
The nonzero metric connection coefficients are found to be
\begin{eqnarray}
 \Gamma^z_{zz}=\frac{-1}{R_f
 g_{zz}}(2(g_{zz}-1)+\frac{e^{2\Lambda}}{\alpha^2}(g_{zz}-1)^2),\nonumber\\
 \quad
 \Gamma^z_{\varphi\varphi}=-\frac{R_t^2 e^{-2\Lambda}}{R_f
 g_{zz}},\quad
 \Gamma^{\varphi}_{z\varphi}=\Gamma^{\varphi}_{\varphi z}=\frac{R_t^2
 e^{-2\Lambda}}{R_f g_{\varphi\varphi}}.
 \end{eqnarray}
The zweibeins are $ e^1_z=\sqrt{g_{zz}},\, e^2_{\varphi}=\rho(z)$,
and the spin connection coefficients are
$\omega_\varphi^{12}=-\omega_\varphi^{21}=\alpha R_t
e^{-2\Lambda}/(\rho(z)\sqrt{g_{zz}})$, so that (\ref{eq:5a})
include the spin connection term
\begin{equation}
\Omega_\varphi=\frac{i\sigma_3\alpha R_t
e^{-2\Lambda}}{2\rho(z)\sqrt{g_{zz}}}.
\end{equation}
Notice that for $-R_f<z<0$ the fields are (\ref{sfield}), and for
$z>0$ (\ref{fields}). After the substitution $
\psi=\left(%
\begin{array}{c}
  u \\ v
\end{array}\right)%
e^{ij\varphi}
$
the Dirac equation (\ref{eq:5a}) takes the form
\begin{eqnarray}
-i(\frac{\partial_z}{\sqrt{g_{zz}}}+\frac{1}{\rho(z)}(j-\frac{\alpha
R_t
e^{-2\Lambda}}{\rho(z)\sqrt{g_{zz}}}-W_{\varphi}-a_{\varphi}))v=Eu,\nonumber\\
-i(\frac{\partial_z}{\sqrt{g_{zz}}}-\frac{1}{\rho(z)}(j+\frac{\alpha
R_t
  e^{-2\Lambda}}{\rho(z)\sqrt{g_{zz}}}-W_{\varphi}-a_{\varphi}))u=Ev.
  \label{uveqn}
\end{eqnarray}
Notice that the momentum $j$ takes half-integer values and enters (\ref{uveqn})
only by the combination $j-a_\varphi$. We expect that in the tube region our model
should be identical to that proposed in~\cite{meletube}. Indeed,
adding $\pm 3/2$ to both $j$ and $a_\varphi$, one obtains in (\ref{fields})
$\int_0^{2\pi} a_\varphi d\varphi=\pm 2\pi M_2/3$. Exactly such circulation was
suggested in~\cite{meletube} for the effective vector potential.

For large $z$ one has
$W_\varphi=0,\; a_\varphi = M_2/3,\sqrt{g_{zz}} \rightarrow 1,\rho(z) \rightarrow
R_t,\Omega_\varphi \rightarrow 0$. In this case, (\ref{uveqn}) takes a similar
to~\cite{meletube} form $-i\sigma_1\partial_z \psi+\sigma_2 m \psi=E\psi$, where
$E=\pm\sqrt{m^2+k^2}$ and the "mass" term $m=-(j-M_2/3)/R_t$.
In this case, the general solution to (\ref{uveqn}) is written as
\begin{eqnarray}
\psi_k =
C\left(%
\begin{array}{c}
  1 \\
  (k-im)/E
\end{array}\right)%
e^{ikz} ,\label{1ddirac}
\end{eqnarray}
with $C$ being a normalization constant. This plane wave solution
has nonzero current $j_z=\bar{\psi}\sigma_1\psi$ which is valid
for the infinite tube. In our case, however, the tube is capped
and the quasi-particles are not able to leave the tube. For this
reason, the current has to be zero everywhere. Taking the
condition $j_z=0$ into account we obtain from (\ref{1ddirac}) the
standing wave solution ($\psi_0=\psi_k+\psi_{-k}$) with zero
current
\begin{eqnarray}
\psi_0 =2C
\left(%
\begin{array}{c}
  \cos (kz+\phi) \\
  i(k \sin (kz+\phi)-m\cos (kz+\phi)\,)/E
\end{array}\right)%
\label{zeroj}.
\end{eqnarray}

Let us analyze the cap region.  For small $\Lambda$ one has
$\rho=R_t\sqrt{2\Lambda},\sqrt{g_{zz}}=\alpha/\sqrt{2\Lambda},W_\varphi=-1,a_\varphi=\pm
3/2$ and the equations (\ref{uveqn}) are written as
\begin{eqnarray}
-i(\sqrt{2\Lambda}\partial_\Lambda+\frac{1}{\sqrt{2\Lambda}}(j-B-1/2))v=\varepsilon
u,\nonumber \\
-i(\sqrt{2\Lambda}\partial_\Lambda-\frac{1}{\sqrt{2\Lambda}}(j-B+1/2))u=\varepsilon
v,\label{spherass}
\end{eqnarray}
with $ B=-1\pm 3/2,\; \varepsilon=E R_t$. Notice that
(\ref{spherass}) are similar to the equations for spherical
fullerene~\cite{kolesn_epjb}. At $\Lambda\approx 0$ the general
solutions (for both $u$ and $v$) have a power-like form $C_1
\Lambda^\alpha+C_2 \Lambda^{-\alpha},\; \alpha\geq 1/2$. Since the
second term cannot be normalized, one has to put $C_2=0$.

\begin{figure}[ht]
\begin{center}
\epsfysize=12cm \epsffile{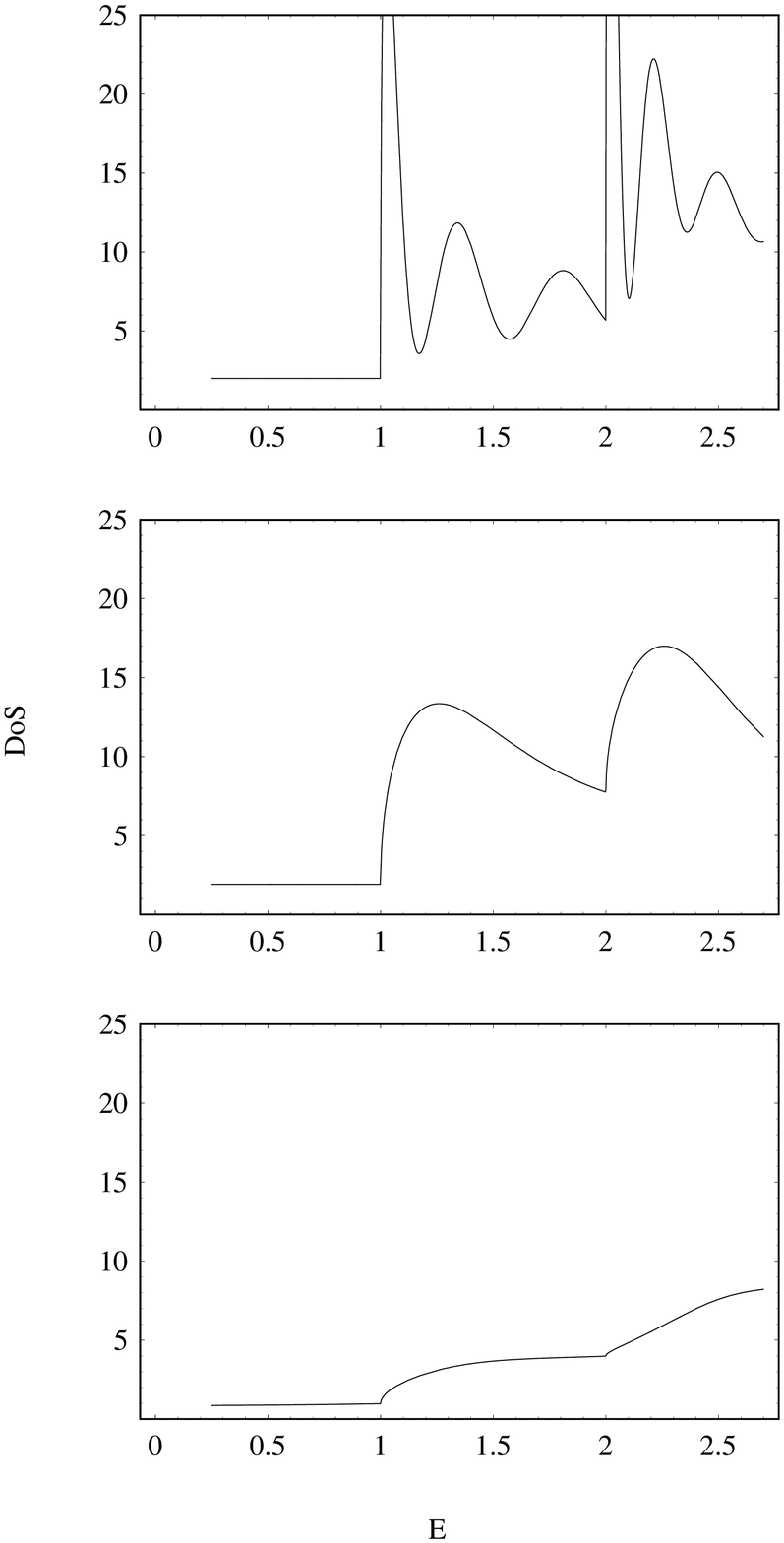} \epsfysize=12cm
\epsffile{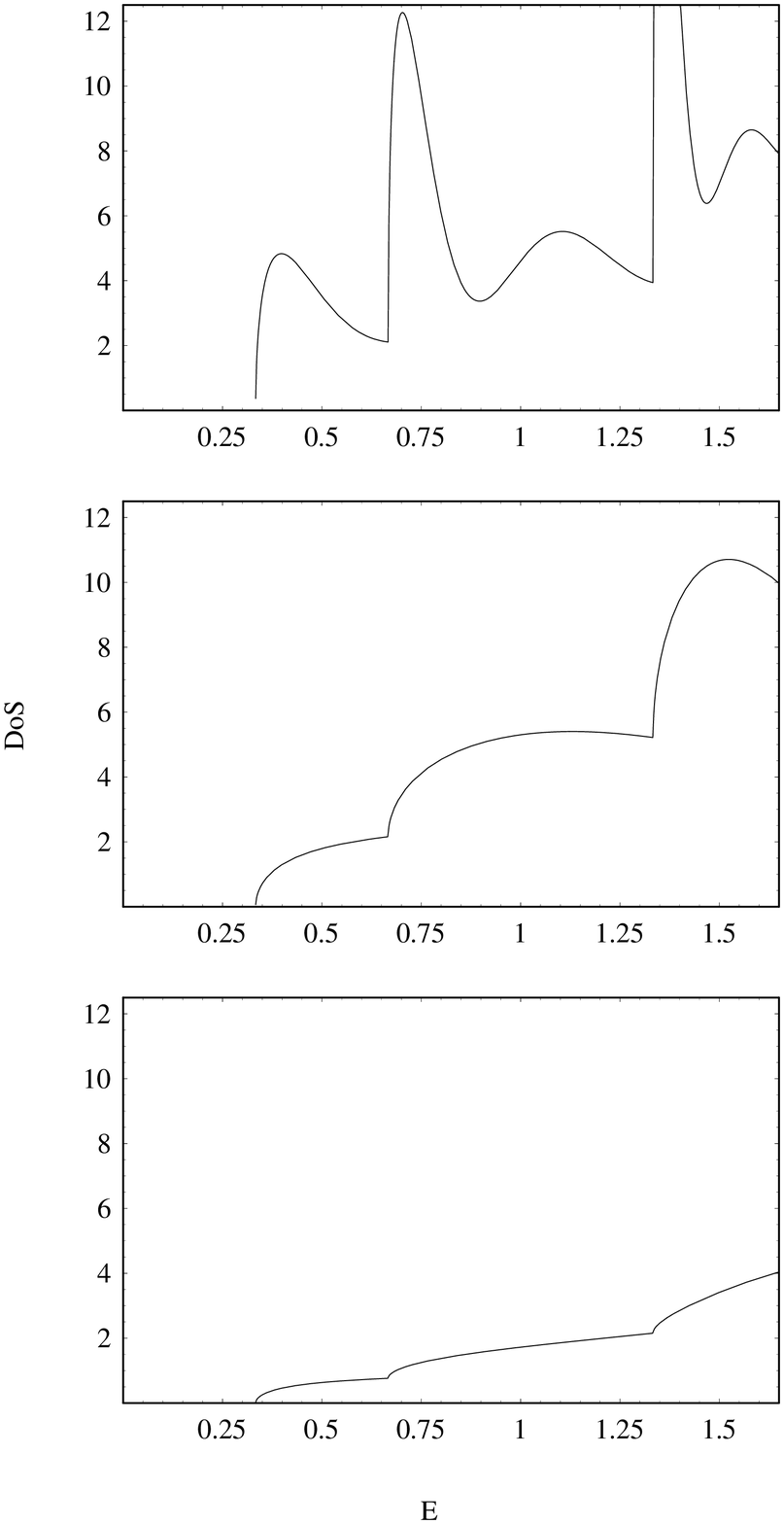} \caption[]{The density of states (per unit
area, in arbitrary units) as a function of the energy in the cap
region (bottom), near the cap (middle) and far from the cap (top).
The case of metallic (left) and semiconducting (right) tubes is
presented. We take $1/\alpha$=0.9, and the energy $E$ is measured
in the units of $\hbar V_F/R_t$.}\label{fig2}
\end{center}
\end{figure}

In the general case, the system (\ref{uveqn}) can be solved
numerically. To this end, the initial conditions are taken to be
(\ref{zeroj}) for $z>>R_f$, and the convergence of the
wavefunction at $z=-R_f$ is obtained by a variation of the phase
$\phi$ up to the moment when $C_2$ turns out to be zero within
calculation accuracy. The wavefunction is
properly normalized, which means $C=\sqrt{DoS_0(E)}$ where
$DoS_0(E)=(\partial E/\partial k )^{-1}$ is the density of states
in the infinite tube. The results of numerical calculations are
presented in figs.\ref{fig1} and \ref{fig2}.
Fig.~\ref{fig2} shows the density of electronic states as a
function of the energy in three regions: far from the cap, near
the cap and on the cap for both metallic and semiconducting cases.
As is seen, the peaks appear at the energies higher than the
threshold energy $m$, where $m R_t=0,\pm 1,\pm 2,..$ for metallic and $m
R_t=\pm 1/3,\pm 2/3,\pm 4/3,..$ for semiconducting case. Notice the
appearance of secondary (less pronounced) peaks far from the cap.
For metallic tube, the constant DoS below the threshold energy is
found.
\begin{figure}[ht]
\begin{center}
\epsfysize=7cm \epsffile{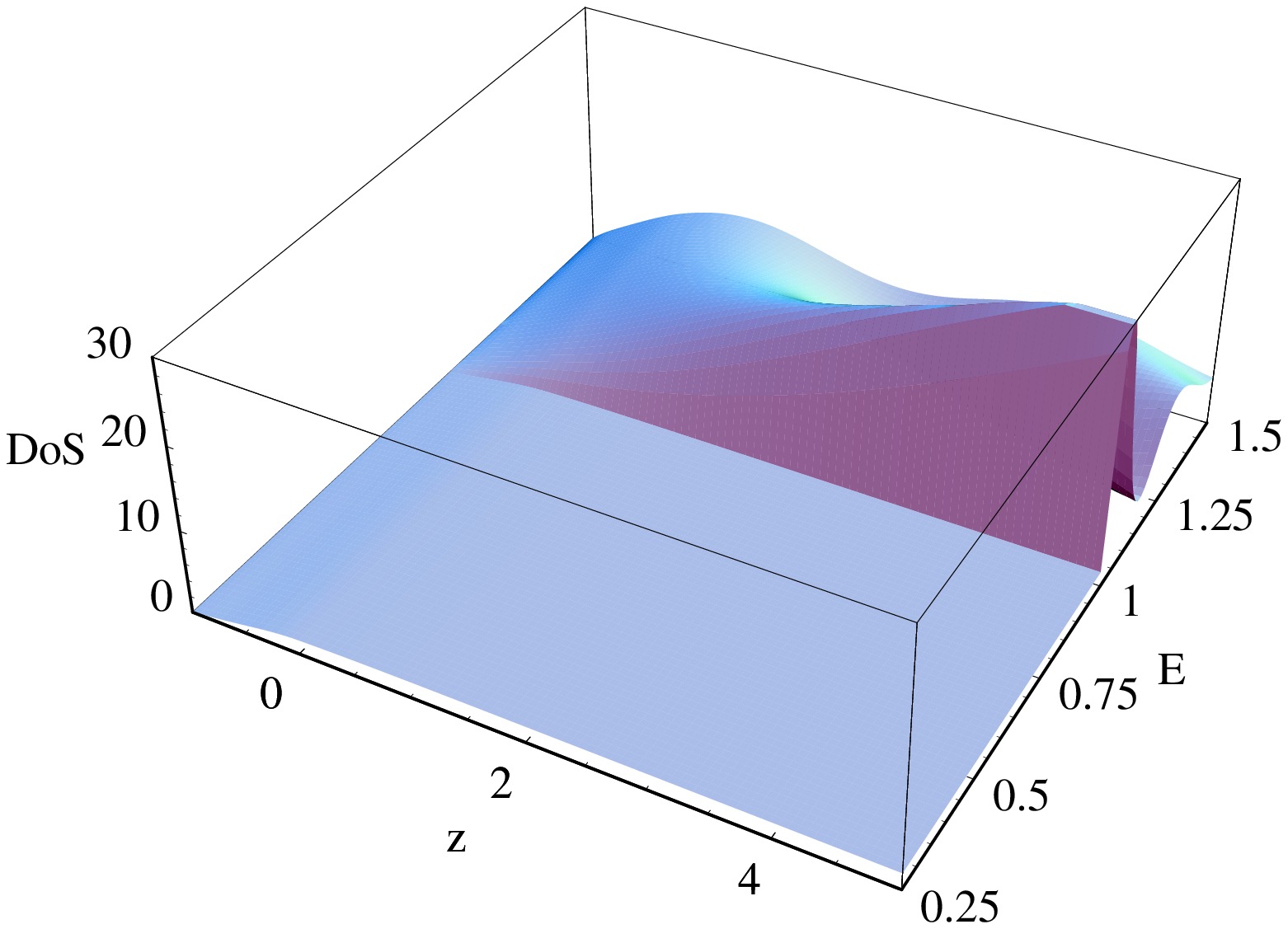} \epsfysize=7cm
\epsffile{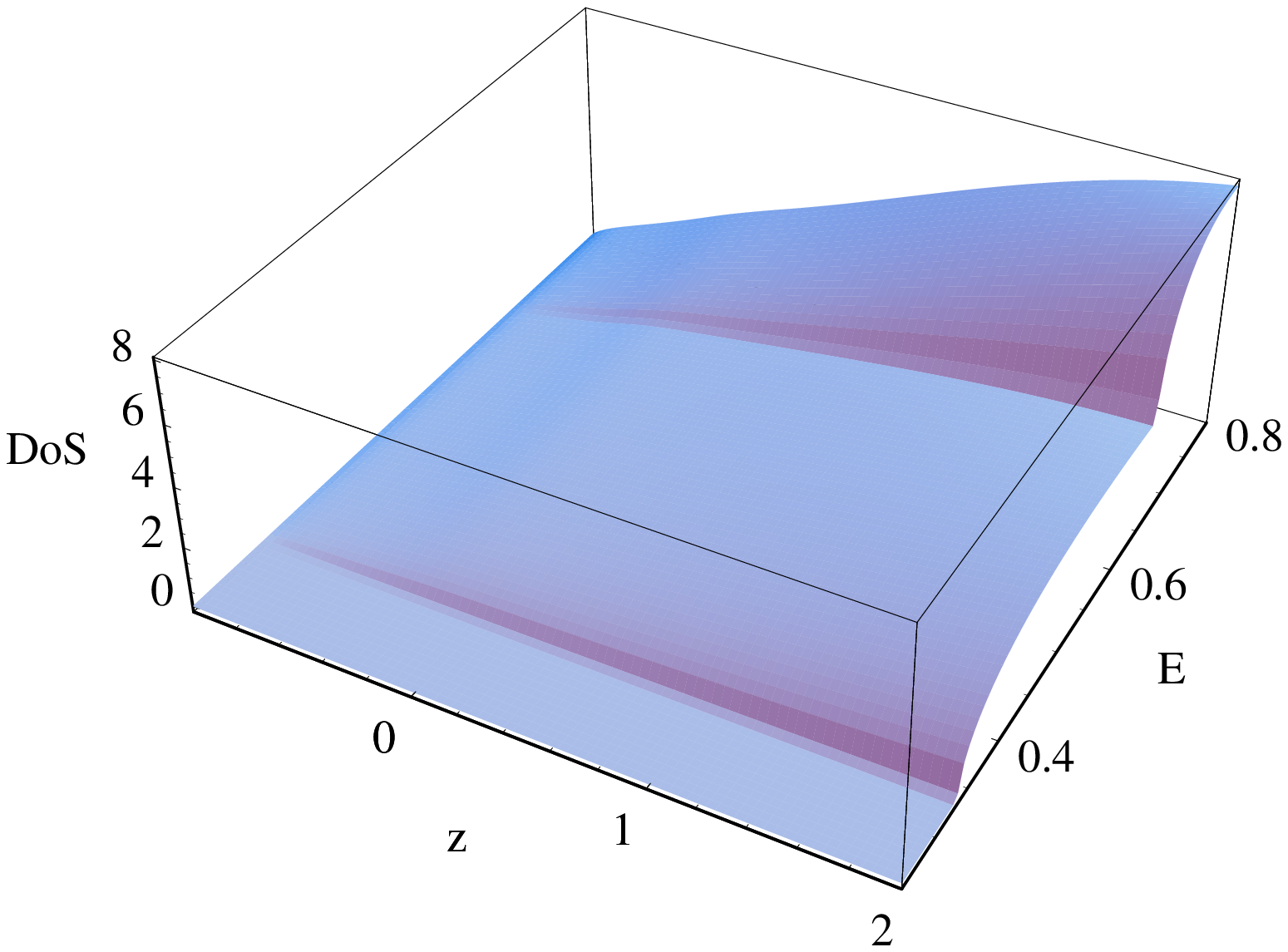} \caption[]{The density of states (per unit
area, in arbitrary units) as a function of the energy (in units of
$\hbar V_F/R_t$) and the coordinate (in units of $R_t$).}
\label{fig3}
\end{center}
\end{figure}
Fig.~\ref{fig3} shows the DoS as a function of the energy and the coordinate.
In the cap region, both the peak values and the constant DoS (in metallic nanotubes)
decrease markedly. At high energies, the harmonic dependence from the coordinate
is seen in fig.\ref{fig3} (left), which is in accordance with (\ref{zeroj}).

The most interesting finding is the finite values of the DoS near
points $E=m$ where the DoS behaves linearly in $E-m$. Moreover, as
is seen from figs.~\ref{fig2},~\ref{fig3}, the smoothed VHS peaks
are shifted to higher energies. Within our model the
dispersion relation has the form $E=\sqrt{k^2+m^2}$ and,
therefore, $DoS_0(E)$ should diverge like $(E-m)^{-1/2}$ when the
energy approaches the threshold $m$. One could
expect a similar behavior in numerical calculations but this was
not observed. It should be noted that the only way to avoid the
singularity is to put $\phi=\pm \pi/2$ in (\ref{zeroj}). In this
case both spinor components turn out to be finite at $E\approx m$
(both $u$ and $v$ become proportional to $k$). Thus, the numerical
results clearly indicate that the phase is fixed
\begin{equation}
\lim_{E \to m+0}\phi(E) =\pm \pi/2.
\label{phase}
\end{equation}
Let us proof this analytically.
Approaching to the cap, the tube tends to decrease its radius
which results in a corresponding increase of the effective mass.
Using the substitution
$\psi=f(\Lambda)\widetilde{\psi}$ where $f(\Lambda)=\exp({-\alpha
e^{-2\Lambda}/2})$ and
expanding the equations (\ref{uveqn}) at large $\Lambda$
up to the order of $e^{-2\Lambda}$ one obtains
\begin{eqnarray}
-i(\partial_\Lambda \mp\mu(1+e^{-2\Lambda}/2))\left(%
\begin{array}{c}\widetilde u\\ \widetilde v\end{array} \right) =\epsilon\left(%
\begin{array}{c}\widetilde u\\ \widetilde v\end{array}
\right),\label{uvbar}
\end{eqnarray}
where $\epsilon=E R_t/\alpha,\  \mu=(M_2/3-j)/\alpha$. This system
is easily reduced to two decoupled equations of the second order
\begin{eqnarray}
\partial_\Lambda^2 \Psi+(A e^{-2\Lambda}+\kappa^2)\Psi=0,
\label{second}
\end{eqnarray}
where $\kappa^2=\epsilon^2-\mu^2$, $A=-(\mu^2\pm\mu)$ where the sign plus (minus)
relates to $\Psi=\widetilde{u}\; (\widetilde{v})$, respectively. In view of
$M_2=0,\pm 1/3,\;j=\pm 1,\pm 2,..$ and $\alpha\approx 1$ the constants
$A$ for $\widetilde u$ and $\widetilde v$ always take opposite signs.
For positive $A$ (which, for example, relates to
$\widetilde{v}$), the substitution $x=\sqrt{A_v}e^{-\Lambda}$ in
(\ref{second}) leads to the Bessel equation of index $i\kappa$
\begin{eqnarray}
x^2 \widetilde{v}''+x\widetilde{v}' +(x^2+\kappa^2)
\widetilde{v}=0,
\end{eqnarray}
and for $\widetilde{u}$ there appears the modified Bessel equation
(with the argument $x=i\sqrt{|A_u|}e^{-\Lambda}$). The solutions
should not diverge at $z\rightarrow\infty\; (x\rightarrow 0)$, so
that they are the Bessel and modified Bessel functions of the first kind,
respectively. For small $k$ (but $kz\gg 1$) one has in the leading order
\begin{eqnarray}
\widetilde{v}\approx iC_v \kappa \log x= iC_v
kz,\nonumber\\
 \widetilde{u}\approx iC_u
\kappa \log x= iC_u kz.
\end{eqnarray}
Substituting this into (\ref{uvbar}) one obtains
the exact relation $iC_v=C_u$ or $iv=u$ in (\ref{zeroj}),
which determines the limit of the phase at $E\rightarrow m$ explicitly
as $\phi=\pi/2$. Assuming a smooth dependence of
the phase $\phi$ on the energy in (\ref{zeroj}) we obtain
both the high peak at $E>m$ and the linear dependence of the DoS near $E=m$.
The amplitude of the peak is found to grow with increasing $z$ (cf. fig.~\ref{fig3}).


In conclusion, we have developed a phenomenological model for the
description of electronic states in capped nanotubes, both
semiconducting and metallic. The influence of pentagons in the cap
region was taken into account via the fictitious compensating
gauge fields. We have suggested a single manifold for the description
of both the nanotube and the cap region. It should be noted that
a similar approach was recently used in~\cite{vozm_epl}
for the description of both curved and defect-free regions in graphene.
The asymptotical solutions in the tube region were
used to obtain the initial conditions for the numerical
calculations of the density of states. The numerical calculations
show that the VHS is smoothed out in DoS. This is explained by the
fact that the phase of the wavefunction in the tube region is
determined by the asymptotic solution near the cap. As it was
analytically shown, in the cap region the mass factor
in~(\ref{uvbar}) markedly increases due to geometry. This results
in fixation of the wave-function phase which, in turn,
provides the smoothing of VHS. Thus, the effect of VHS smoothing
is geometrical in its origin within our approach.

The numerical calculations show the finite peaks at energies
greater than the threshold energies. The amplitude of the peaks
grows with $z$. At threshold points $E=m$, the DoS is found to be
linear in $E-m$ rather than manifests the standard singular
behaviour. The obtained results are in qualitative agreement with
the experimental measurements and tight-binding calculations for
the closed nanotube performed in~\cite{kim}. At the same time,
near the cap an increase of DoS was observed and explained
in~\cite{kim} by the specific distribution of six pentagons on the
non-spherical cap. Notice that in our model the defects in the cap
region are taken into account by effective continuous fields. For
this reason, the localized states are smeared out and there are no
pronounced peaks. Between the peak energies the DoS on the cap was
found to be reduced in comparison with the tube region~\cite{kim}.
In our model the decrease of the DoS on the cap is found at all
energies.

\acknowledgments
This work has been supported by the Russian
Foundation for Basic Research under grant No. 05-02-17721.

\end{document}